\documentclass[journal=prb,manuscript=article, twocolumn]{revtex4-2}
\usepackage{amsmath}
\usepackage{amssymb}
\usepackage{graphicx}
\usepackage{bm}
\usepackage{hyperref}
\usepackage{xcolor}
\usepackage{subfigure}

\begin{document}

\title{Quantum Electrodynamics in an Electrolyte Medium Driving Entanglement Between Graphene Sheets}

\author{David A. Miranda}
\email{dalemir@uis.edu.co}
\affiliation{Universidad Industrial de Santander, Cra 27 Cll 9, 680002, Bucaramanga, Colombia}

\author{Edgar F. Pinzón}
\affiliation{Department of Engineering, Physics and Mathematics, São Paulo State University (UNESP), CP 355, Araraquara, São Paulo, 14800-060, Brazil}

\author{Paulo R. Bueno}
\email{paulo-roberto.bueno@unesp.br}
\affiliation{Department of Engineering, Physics and Mathematics, São Paulo State University (UNESP), CP 355, Araraquara, São Paulo, 14800-060, Brazil}

\date{\today}

\begin{abstract}
The quantum entanglement phenomenon was demonstrated to operate on a bipartite entangled system composed of two single layers of graphene embedded in an electrolytic medium (which did not permit the transport of electrons) and subjected to an external time-dependent electric perturbation driven by a potentiostat equipped with a frequency response analyser. Time-dependent perturbation-mediating entanglement was hypothesised because of the equivalent quantum resistive-capacitive circuit frequency of each single-layer graphene system that obeys quantum electrodynamics principles. \textit{De facto}, quantum electrodynamics, associated with the massless fermionic characteristics in graphene sheets, was observed at room temperature and electrolyte medium under a time-dependent modulation, and the entanglement between the two sheets is consistent with a Hilbertian subspace mathematical examination of the phenomenon. Remarkably, this experiment, among numerous other quantum electrochemical experiments conducted by this research group that follow quantum electrodynamics principles, highlights the generality of the methodology for studying entanglement phenomena, leading to alternative methods for investigating Weyl semi-metal structures and designing room-temperature quantum applications.

\end{abstract}

\maketitle

\section{Introduction}\label{sec:introduction}

Entanglement between fermions is a quantum mechanical phenomenon without analogous classical mechanical particles or waves. This phenomenon arises because of the interplay between two or more Hilbert subspaces in which a complete description cannot be attained by considering separable states within each subspace. A particular description (generally mathematical) of entangled states is required to account for quantum entanglement. 

The entangled phenomenon was initially postulated by Einstein, Podolsky, and Rosen (EPR)~\cite{EPR1935} as a means of questioning the completeness of the quantum mechanical theory, leading to what is referred to as the EPR paradox. According to Einstein's interpretation of quantum mechanics, the apparent nonlocal interactions arising from quantum entanglement suggest the presence of hidden variables, rendering the theory incomplete. To address this issue, Bell formulated a theoretical framework known as the Bell inequalities~\cite{Bell1964} to substantiate the EPR paradox. In subsequent years, Clauser et al.~\cite{Clauser1969} devised an experiment to test the validity of Bell inequalities, and Aspect ~\cite{Aspect1982a, Aspect1982b} conducted groundbreaking experiments that demonstrated the violation of Bell inequalities, thereby suggesting the completeness of quantum mechanics. These experiments demonstrated that the entanglement phenomenon predicted by quantum mechanics is an inherent feature of the quantum world. It was only in 1989 that Werner proposed the first mathematical definition of the quantum entanglement phenomenon~\cite{Werner1989}. More recently, in 2013, Balachandran et al.~\cite{Balachandran2013} proposed a theoretical framework to describe quantum entanglement in systems comprising identical particles, further advancing our understanding of this phenomenon.

Quantum entanglement has been investigated for a wide range of material systems, including trapped ions \cite{Leibfried2005, Hffner2005, Riebe2008, Bock2018, Kobel2021}, Josephson junctions \cite{Peugeot2021, Huang2022}, photons \cite{Bouwmeester1997, Rab2017}, and electron spins \cite{Schaibley2013}. This study proposes an alternative approach that uses an electrochemical setup consisting of two entangled singly-layer graphene (SLG) systems to form a bipartite system. The two SLGs are spatially separated by an electrolyte environment, which is not a conductor of electrons, owing to the absence of a redox probe.

The role of the electrolyte in the experiment was solely to permit screening of the electric field within SLGs without permitting the transport of electrons through the electrolyte, thus keeping the two SLGs electronically and spatially separated. The electric-field screening of the electrons in SLGs embedded into an electrolyte permits the access of quantum electrodynamic information within SLG systems~\citep{lopes2021measuring, Bueno-2022} and demonstrates that the quantum electrodynamics of the two spatially separated and electrically isolated systems are quantum mechanically entangled because they possess equivalent quantum mechanical energy $E$ states and frequency dynamics $\nu$ that obey Planck--Einstein electrodynamics with $\nu = E/h$, where $h$ is the Planck constant.

Accordingly, by measuring the quantum electrodynamics of graphene at room temperature in an electrolyte environment, quantum mechanical entanglement was established between two spatially separated and electrically isolated ~\footnote{Electrons are not allowed to diffuse or migrate between the two SLG foils.} SLG systems. The experiment allowed us to measure the characteristic frequency $\nu = E/h$ of one SLG system and demonstrate that it entangles with another. Here, $\nu$ was determined using a time-dependent method referred to as impedance-derived capacitance spectroscopy, in which a complex capacitance function of type

\begin{equation}
 \label{eq:Complex-Cq}
	C^{*}(\omega) = \frac{C_{q}}{1 + j\omega \tau} \sim C_{q} \left(1 - j\omega \tau \right)
\end{equation}

\noindent is obtained~\cite{bueno2023quantum, Bueno-book-2018, lopes2021measuring, Bueno-2022}. The term $C_{q}$ corresponds to the equilibrium quantum capacitance measured for $\omega \rightarrow 0$, where $\omega$ is the angular frequency and $j = \sqrt{-1}$ is an imaginary number. It is noteworthy that $\tau = R_q C_q$ in Eq. ~\ref{eq:Complex-Cq} corresponds to a characteristic quantum resistive-capacitive (RC) time constant, where $R_q = 1/G \propto h/e^2$ and $C_q$ is associated with the density of states $(dn/dE) = C_q/e^2$ (DOS), leading to quantum electrodynamics, which can be summarised in terms of a series combination of $R_q$ and $C_q$~\cite{Bueno-2020-ET} quantum circuit elements of the system. 

This quantum RC dynamics is directly related to the characteristic frequency of the system $\nu = 1/\tau$ defined within the quantum rate theory~\cite{Bueno-2020-ET, bueno2023quantum} as the ratio between the reciprocal of the von Klitzing constant~\cite{von-klitzing-1980} ($h/e^2$) and $C_q$, in which $\nu$ is defined as $\nu = 1/\tau = E/h \propto e^2/hC_q$, obeying quantum electrodynamics~\citep{Dirac-1928} where $E = e^2/C_q$ is the energy per particle~\footnote{Note that $dE/dn = e^2/C_q$ and when the single-particle dynamics $dn = 1$ are settled, $e^2/C_q$ is the energy of a single ground state level.} associated with the quantum-mechanical state of the system.

The time-dependent quantum electrodynamic phenomena in Eq. ~\ref{eq:Complex-Cq} not only allowed us to investigate the quantum entanglement between an SLG bipartite system, as will be demonstrated here, but also showed us that this phenomenon governs the dynamics of electrochemical reactions~\citep{bueno2023quantum, alarcon2021perspective}, allowing us to better understand the mechanisms of respiration~\citep{Bueno-2024-bioreview}. Furthermore, the quantum-rate phenomenon has allowed the development of electroanalytical spectroscopic methods for measuring the electronic structure of graphene~\citep{Lopes-2024-SLG-structure} and quantum dots~\citep{Pinzon-2024-QDs}, in addition to characterising the nanoscale dynamics of semiconducting molecular electronic junctions~\citep{Pinzon-2024-Meletronics} and fabricating biological sensors with quantum resolution~\citep{Garrote-2020-Nature}.

In other words, quantum rate theory establishes a foundation that permits the unification of electrochemistry and nanoscale electronics~\citep{Bueno-Davis-2020}. This phenomenon occurs because in an electrochemical environment, the energy $E = e^2/C_q$ follows the quantum electrodynamics phenomenon and is thus governed by $E/h = e^2/hC_q$, but with a frequency that is modulated by the dynamics of the solvent, constituting a low-energy quantum electrodynamics phenomenon. This low-frequency electrodynamic phenomenon results from the intertwined electrostatic and quantum energies. The electrochemical energy $e^2/C_\mu$ corresponds to the contribution of the capacitance of the electrolyte $C_e$ in a serial combination with $C_q$ such as $1/C_{\mu} = 1/C_e + 1/C_q$. Therefore, the presence of cations and anions in the electrolyte permits a suitable screening of the electric field associated with perturbing quantum states, permitting $C_e$ to be of the same order of magnitude as $C_q$ (or vice-versa) such as $C_e \sim C_q$, leading to $e^2/C_{\mu} = e^2/C_e + e^2/C_q = 2e^2/C_q$ as $C_e \sim C_q$. In other words, because of energy degeneracy denoted as $g_e = 2$ both $e^2/C_e \sim e^2/C_q$~\citep{bueno2023quantum} are accessible.

This energy degeneracy of the system has been experimentally demonstrated~\cite{alarcon2021perspective, bueno2023quantum, sanchez2022quantum} for different systems measured in appropriate electrolyte environments, including graphene~\citep{Bueno-2022, Lopes-2024-SLG-structure, lopes2021measuring}. Additionally, if the spin degeneracy associated with the conductance quantum $e^2/h$ is accounted for, the total energy degeneracy of the states is $E = g_s g_e (e^2/C_q)$, where $g_s$ is the electron spin degeneracy contribution and $g_e$ is associated with the energy degeneracy owing to the electric-field screening of the electrolyte over measurable quantum states.

The impedance-derived electrochemical measurements and analysis of $C_q$ provided by Eq.~\ref{eq:Complex-Cq}, aligned with the premises of quantum-rate (QR) theory ~\citep{bueno2023quantum,Bueno-2020-ET,Bueno-2022}, has led to significant advances in our understanding of electrodynamics at the nanoscale within an electrolyte environment, as noted above. Considering that $E = e^2/C_q$ is the energy intrinsically associated with the electronic structure, this implies the Planck--Einstein relationship in which $E = h\nu = \hbar \textbf{c}_* \cdot \textbf{k}$ follows a linear relationship dispersion between the energy $E$ and the particle wave vector, where $\textbf{c}_*$ is the Fermi velocity and $\hbar$ is $h$ divided by $2\pi$. Therefore, the QR analysis of quantum electrodynamics within an electrolyte environment complies with Dirac's~\citep{Dirac-1928} relativistic wave mechanics, in which electrons with null rest mass dynamics are referred to as massless fermionic particles, complying with the relativistic equation $E^2 = \textbf{p}^2\textbf{c}_*^2 + m_0^2 c_*^4$, where $E$ is the total energy and $m_0$ is the resting mass.

In summary, QR premises within the superposition of electrostatic and quantum electronic energies intrinsically imply, in physics terminology, low-energy quantum electrodynamics, in which low energy is due to electrodynamics within an electrolyte environment. This analysis has led to a significant increasing in the understanding of quantum electrochemistry~\citep{bueno2023quantum}. In particular, the implementation of QR theory to graphene has not only permitted us to measure its electronic structure~\citep{Lopes-2024-SLG-structure} but has also led to the possibility of studying conductive and capacitive V-shapes~\cite{lopes2021measuring} separately or in combination. 

The advantage of using an electrolyte environment is that it permits the study of quantum effects at room temperature and \textit{in situ}~\cite{bueno2023quantum, Bueno-2022} using electric impedance analysis, opening up possibilities for investigating routine quantum phenomena such as entanglement, which would otherwise require low-temperature and high-vacuum conditions, requiring a complex and expensive experimental apparatus. Conversely, the measurements associated with quantum electrodynamics determined by Eq. ~\ref{eq:Complex-Cq} may be conducted using an inexpensive handheld or benchtop potentiostat, permitting the setting of three-electrode measurements, which are referred to within electrochemical nomenclature as the working (WE), counter (CE), and reference (RE) electrodes, all connected to a potentiostat equipped with a frequency response analyser (FRA), as a requirement for performing time-dependent measurements that comply with Eq. ~\ref{eq:Complex-Cq}.

For example, by using the FRA module of a potentiostat, harmonic electric potential perturbations $\tilde{V}(t) = V_0\exp(j\omega t)$ can be applied with a small amplitude $V_0$ over a stationary $\bar{V}$ potential established between the WE and RE, resulting in a time-dependent $V(t) = \bar{V} + \tilde{V}(t)$ modulation over the stationary $\bar{V}$ potential. The response to $\tilde{V}(t)$ is a time-dependent electric current over a stationary electric current $\bar{i}$ owing to $\bar{V}$, that is $i(t) = \bar{i} + \tilde{i}(t)$, where $\tilde{i} = I_0 \exp(j\omega t - \phi)$ is the modulated electric current response measured between the CE and RE, where $\phi$ states for the phase difference established between the potential perturbation and the electric current response. No electric current was imposed on the RE, which acts as a non-polarizable electrode. Accordingly, for each $\omega$, the electrical impedance is determined to be $Z = (V_0/I_0) \exp(j\phi)$, from which $C^*(\omega)$ in Eq. ~\ref{eq:Complex-Cq} is $C^*(\omega) = I_0/(j\omega V_0) \exp(-j\phi)$.

The underlying hypothesis of the present study is to demonstrate whether bipartite quantum entanglement between two SLGs is attainable. This hypothesis was experimentally confirmed by noting that whenever the WE and CE electrodes were in contact with different but equivalent SLGs, an entanglement phenomenon was observed and measurable between the two SLGs. We anticipate that the observation of the quantum entanglement phenomenon introduces an unexpected arrangement of the SLG's capacitance in which the SLG capacitance of the WE is approximately half of its initial value, thus evidencing the entanglement phenomenon between the WE and CE quantum energy levels. This correlation is not present or cannot be measured by the potentiostat if classical circuit elements are in contact with the CE, as further discussed below.

The time-dependent perturbation and quantum mechanical rate dynamics $\nu = E/h$ associated with a quantum RC circuit dynamic description of the phenomenon permit the entanglement of two equivalent SLGs, which is consistent with a bipartite (mathematical) entangled analysis of the problem. The experiments conducted in this study consisted of three main steps. The \textit{first} (section~\ref{sec:Operation-Potentiostat}) aims to clarify and attest to the operation of a potentiostat within an FRA instrument using the classical setting of circuit elements in contact with the WE and CE, with the measurements conducted in three- and two-electrode settings, to demonstrate and confirm that, the impedance measured in the WE cannot be affected by contact with the CE in the three-electrode setting. In the \textit{second} (section~\ref{sec:measuring-SLGs-entanglement}) step, the influence of the impedance of the CE on the WE was investigated when the quantum circuit elements were in contact with both the WE and CE, in which the WE and CE were in contact with equivalent graphene sheets within the quantum RC circuit dynamics. Finally, the \textit{third} (Section ~\ref{sec:entanglement-biosensing}) step involves examining how the WE is affected of changes in the impedance of an entangled CE (modified with a receptor) by an \textit{in situ} interaction of the CE considering a specific biological target that can be captured by the attached receptor.

\section{Materials and Methods}
\label{sec:methods_materials}

\subsection{Chemical Reagent and Solutions} 

All reagents were of analytical grade and purchased from Sigma-Aldrich. A phosphorus buffer solution (PBS) (pH 7.4) was prepared by dissolving 10 mM Na$_2$HPO$_4\cdot12$ H$_2$O and 2 mM KH$_2$ PH$_4$ in Milli-Q ultrapure water with a resistivity of 18.2 M$\Omega$ to serve as an electrolyte in all experiments reported in the present work. Solutions of N-(3-dimethylaminopropyl)-N'-ethylcarbodiimide (EDC) 400 mmol L$^{-1}$ (Sigma-Aldrich, ca. 03450), and N-hydroxy succinimide (NHS) 100 mmol L$^{-1}$ (Sigma-Aldrich, ca. 130672) were prepared in Milli-Q water. Monoclonal anti-nucleocapsid (anti-N) antibody (ab272852) and rabbit monoclonal antibody [EPR24334-118] specific to the SARS-CoV-2 nucleocapsid protein (ab271180) were purchased from Abcam. The biological assays employed Superblock (ca. 37515) from Thermo Fisher Scientific as a blocking buffer in PBS.

\subsection{Single-Layer Graphene}
\label{sec:SLG-fabrication}

SLG was obtained from Graphenea Company and deposited by chemical vapour deposition (CVD) on a SiO$_2$/Si substrate in which the SiO$_2$ layer thickness was 90 nm. Electric contacts to SLG were made by first depositing a layer of titanium (10 nm) and subsequently gold (100 nm) by radio frequency sputtering using a metal mask (10 mm × 10 mm) with an open slot (8 mm × 2 mm), allowing the deposition of titanium and gold in a specific region while preserving the SLG structure for the measurements.

\begin{figure*}[t]
\centering
\includegraphics[width=16cm]{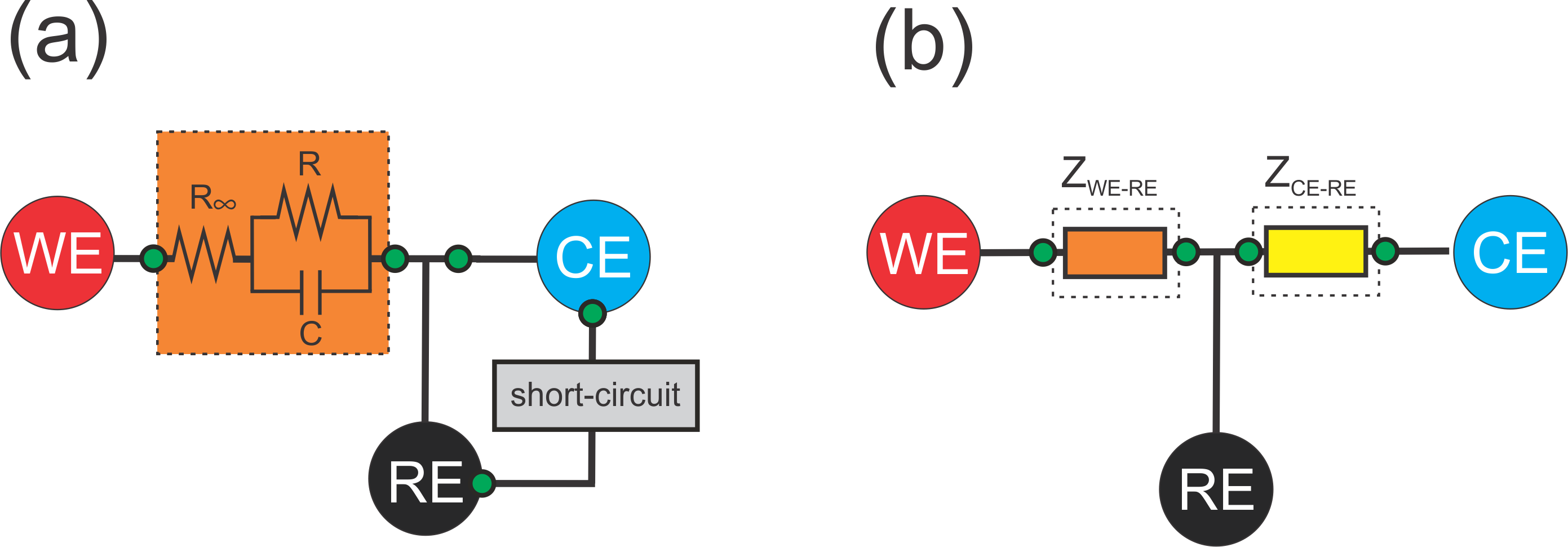}
\caption{Schematic  of two- and three-terminal electrochemical cell setup measured in a potentiostatic mode. Note in (a) there is a $[R_\infty(RC)]$ circuit and the impedance of this circuit can be analysed in two- or three-terminal configurations. In the two-terminal configuration, the analysis is made by short-circuiting CE and RE terminals, in which the measured impedance is the impedance of the total cell comprising the impedance of the combined WE and CE. (b) Typical three-terminal configuration of an electrochemical cell in which the impedance $Z_{WE-RE}$ measured in the WE is separate from the impedance $Z_{CE-RE}$ of the CE, and in a three-terminal system, the impedance measured by the potentiostat is only that of the WE.  Accordingly, the design of the potentiostat ensures that the impedance of the CE does not affect the measurement of the impedance of the WE in a three-terminal configuration, as is experimentally demonstrated in this study considering the classical measurement conditions.}
\label{fig:potentiostate-behavior-setups}
\end{figure*}

\subsection{ Operation of the Potentiostat }
\label{sec:Operation-Potentiostat}

To demonstrate the correct operation of the electrochemical instrumentation (potentiostat) for studying the entanglement phenomenon, three measurements with different configurations of impedance of the terminals of the potentiostat were performed using only the impedance of classical electric circuit elements. This circuit analysis aimed to demonstrate the classical electronic operation of two- and three-terminal potentiostatic measurements in which dummy cell circuit elements were used as impedance connections to the terminals of the potentiostat. An AUTOLAB potentiostat/galvanostat PGSTAT30 equipped with an FRA module was used to measure the electrical impedance spectrum (EIS) in the two- and three-terminal electrochemical modes.

Figure~\ref{fig:potentiostate-behavior-setups}\textit{a} shows the three-electrode terminals of an electrochemical cell measured in the potentiostatic mode of operation in which a classical electric circuit is in contact with the WE electrode (shown in red). This electric circuit corresponds to a $R_\infty$ resistance placed in series `[]' to a parallel `()' arrangement of a resistor $R$ and a capacitor $C$, which is thus denoted as a $[R_\infty(RC)]$ circuit. This system follows the impedance relaxation function shown in Eq.~S1  within the characteristic time constant, $\tau = RC$. Because potentiostats can also operate using either two- or three-terminal settings, Figure~\ref{fig:potentiostate-behavior-setups}\textit{a} demonstrates that measurements in the two-terminal setup are conducted by short-circuiting the RE to the CE.

\begin{figure*}[t]
\centering
\includegraphics[width=18cm]{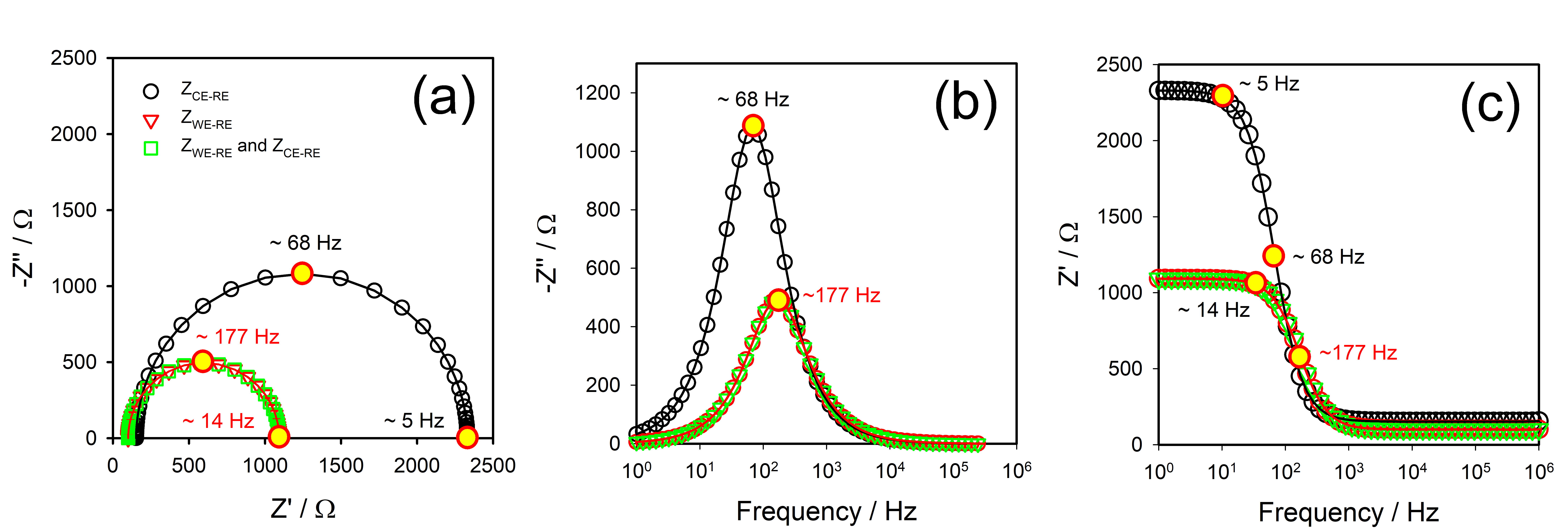}
\caption{(a) Nyquist plot of the $Z_{CE-RE}$ and $Z_{WE-RE}$ impedance, in which $Z_{CE-RE}$ and $Z_{WE-RE}$ were measured separately in a two-terminal setting, as shown in Figure~\ref{fig:potentiostate-behavior-setups}\textit{a}. The plot labelled as \lq\lq $Z_{WE-RE}$ and $Z_{CE-RE}$\rq\rq\ corresponds to the measurement conducted using a three-terminal setting, as illustrated in Figure~\ref{fig:potentiostate-behavior-setups}, in which $Z_{WE-RE}$ corresponds to the impedance connected to the WE and $Z_{CE-RE}$ to the impedance connected to the CE. (b) and (c) depict the same response shown in (a) but using Bode diagrams of imaginary $Z''$ (b) and real $Z'$ (c) components of the complex impedance function $Z^*$ plotted as a function of the frequency. The lines correspond to the data fitting of the equivalent circuit. The results of the fitting are discussed and shown in the SI document and are in good agreement with the graphical analysis of the characteristic frequencies shown in this figure.}
\label{fig:Z1-Z2-Z1Z2}
\end{figure*}

The known $Z_{WE-RE}$ and $Z_{CE-RE}$ circuit impedances shown in Figure~\ref{fig:potentiostate-behavior-setups}\textit{b} were measured separately in a two-terminal setting, as shown in Figure~\ref{fig:potentiostate-behavior-setups}\textit{a}. Each $Z_{WE-RE}$ and $Z_{CE-RE}$ corresponds to the impedance of $[R_\infty(RC)]$ circuits, differing only in the magnitude of the circuit elements. For instance, in a two-terminal configuration of the dummy cell, $Z_{CE-RE}$ corresponded to the impedance of a $[R_\infty(RC)]$ circuit in which the elements were settled as $R_\infty$ = 150 $\Omega$, $R$ = 2.2 k$\Omega$, and $C$ = 1 $\mu$F, whereas $Z_{WE-RE}$ corresponded to the impedance of a $[R_\infty(RC)]$ circuit in which the elements of the circuit were $R_\infty$ = 100 $\Omega$, $R$ = 1.0 k$\Omega$, and $C$ = 1 $\mu$F. Each impedance has a characteristic frequency given by $\nu = 1/(2\pi RC)$ such as that in a two-terminal setting each of them is $\nu_{CE-RE} = 1/[2\pi (2.2 k\Omega) \times (1 \mu F)] \sim$ 72 Hz and $\nu_{WE-RE} = 1/[2\pi (1 k\Omega) \times (1 \mu F)] \sim$ 159 Hz.

After measuring and confirming that the values of each circuit element were accurately measured in a two-configuration setting using Nyquist and Bode plot analysis (Figure~\ref{fig:Z1-Z2-Z1Z2}) or by fitting analysis of the plots to an impedance relaxation function, (see Section S1), $Z_{WE-RE}$ was contacted with the WE and $Z_{CE-RE}$ was contacted with the CE; thus, measurements were conducted in a three-terminal setting. The response is shown by the green square in Figure~\ref{fig:Z1-Z2-Z1Z2}, which is equivalent to that of $Z_{WE-RE}$ (red inverted triangle) measured in the two-terminal setting (see also Section S1), demonstrating that in a three-terminal configuration, the measured impedance corresponds to the impedance of a circuit in contact with the WE, and the impedance of the CE cannot affect measurements conducted in the WE. These results agreed with our expectations and  corresponded to the accurate operation of a three-terminal electrochemical cell configuration measured in the potentiostatic mode.

\subsection{Measuring Quantum Entanglement between SLGs}
\label{sec:measuring-SLGs-entanglement}

Figure~\ref{fig:entanglement-setup} depicts the electrochemical experiments performed at room temperature (298 K). Here, (a) depicts the  impedance measurements performed with the WE in contact with an SLG and CE to a Pt in the format of a foil, while (b) shows the results upon contact with the CE using another SLG in place of the Pt. In both configurations, the RE terminus was Ag$\mid$AgCl (3 M KCl), which served as the reference true chemical potential. The experiments were performed in quadruplicate.

Impedance measurements were in a PBS environment at a stationary potential of -0.13 V, corresponding to the open circuit potential (OCP) (see also Figure~\ref{fig:V-shapes}). The frequency interval of the measurement ranged from 100 kHz to 1 Hz with a peak-to-peak potential amplitude $V_0$ of 10 mV. The real ($C'$) and imaginary ($C''$) components of the complex capacitance function $C^*(\omega)$ were obtained from the raw data of the complex impedance $Z^*(\omega)$ using the relationship $C^{*}(\omega) = 1/[j\omega Z^*(\omega)]$, where the real and imaginary components of $C^*(\omega)$ are obtainable as $C' = \varphi Z''$ and $C'' = \varphi Z'$, where $\varphi = (\omega |Z|^2)^{-1}$ and $\omega$ is the angular frequency of the modulated potential $\tilde{V}(t) = V_0\exp(j\omega t)$. $|Z|$ corresponds to the modulus of the impedance $|Z| = [(Z')^2 + (Z'')^2]^{1/2}$ obtained from the real $Z'$ and imaginary $Z''$ components of the complex impedance function $Z^*(\omega) = Z' - j Z''$.

\subsection{Biological Sensing through the Counter Electrode Entangled with the Working Electrode}
\label{sec:entanglement-biosensing}

The two SLG electrodes obtained as described in Section ~\ref{sec:SLG-fabrication} were modified with a 1 mM solution of 1-pyrenebutyric acid (PA) for 2 h. The modification of SLGs with PA molecules was established through $\pi-\pi$ stacking. Initially, impedance measurements were performed using a PA-modified SLG in contact with the WE and Pt to the CE with Ag$\mid$AgCl serving as the RE, in which only the impedance of the WE was measured in PBS at pH 7.4, as depicted in Figure~\ref{fig:entanglement-setup}\textit{a}. The diagnosis of COVID-19 using the quantum RC properties of graphene in the setting shown in Figure~\ref{fig:entanglement-setup}\textit{a} was previously studied by our research group; more details can be found in ~\citep{Garrote-ACS-2022}.

Subsequently, measurements were recorded by contacting the CE with another PA-modified SLG, thus forming a bipartite entangled system. This was performed to demonstrate that the entangled biological sensing impedance of the CE can be used to detect nucleocapsid protein (N-protein) as biomarker of COVID-19 through the measurement of the impedance of the WE using the settings depicted in Figure~\ref{fig:entanglement-setup}\textit{b}.  In other words, the anti-N-protein-receptive PA-modified SLG surface in contact with the CE terminal of the electrochemical cell was measured in the WE owing to entanglement with the CE.

The chemical modification of the SLG used in the CE terminal of the electrochemical cell was achieved using the carboxylic acid groups of the immobilized PA molecules. This was performed after chemical activation with 0.2 M of EDC and 0.05 M of NHS for 30 min to react with the acid groups to the amine groups present in the anti-N-protein~\citep {Garrote-ACS-2022}. Immobilisation of the anti-N-protein on the surface was achieved by incubating the activated PA-SLG surface in a 1 $\mu$g mL$^{-1}$ protein solution in PBS for 1 h. To prevent nonspecific binding to the remaining carboxylic acid groups of the interface, a super-block buffer was employed with an incubation time of 30 min. Finally, the CE biological sensing terminal of the electrochemical cell was incubated in PBS with 500 ng mL$^{-1}$ of N-protein for 30 min, and the biological antigen-antibody interaction was probed by impedance measurements of an SLG in contact with the WE~\citep{Garrote-ACS-2022}. These experiments were performed in triplicate. Statistical and quantitative analyses are shown in Section SI.3.

\section{Analysis of the Results}
\label{sec:experimental_results}

\subsection{Potentiostat Operating with Quantum RC Circuit Electrodynamics}
\label{sec:potentiostate}

\begin{figure}[th]
\centering
\includegraphics[width=8cm]{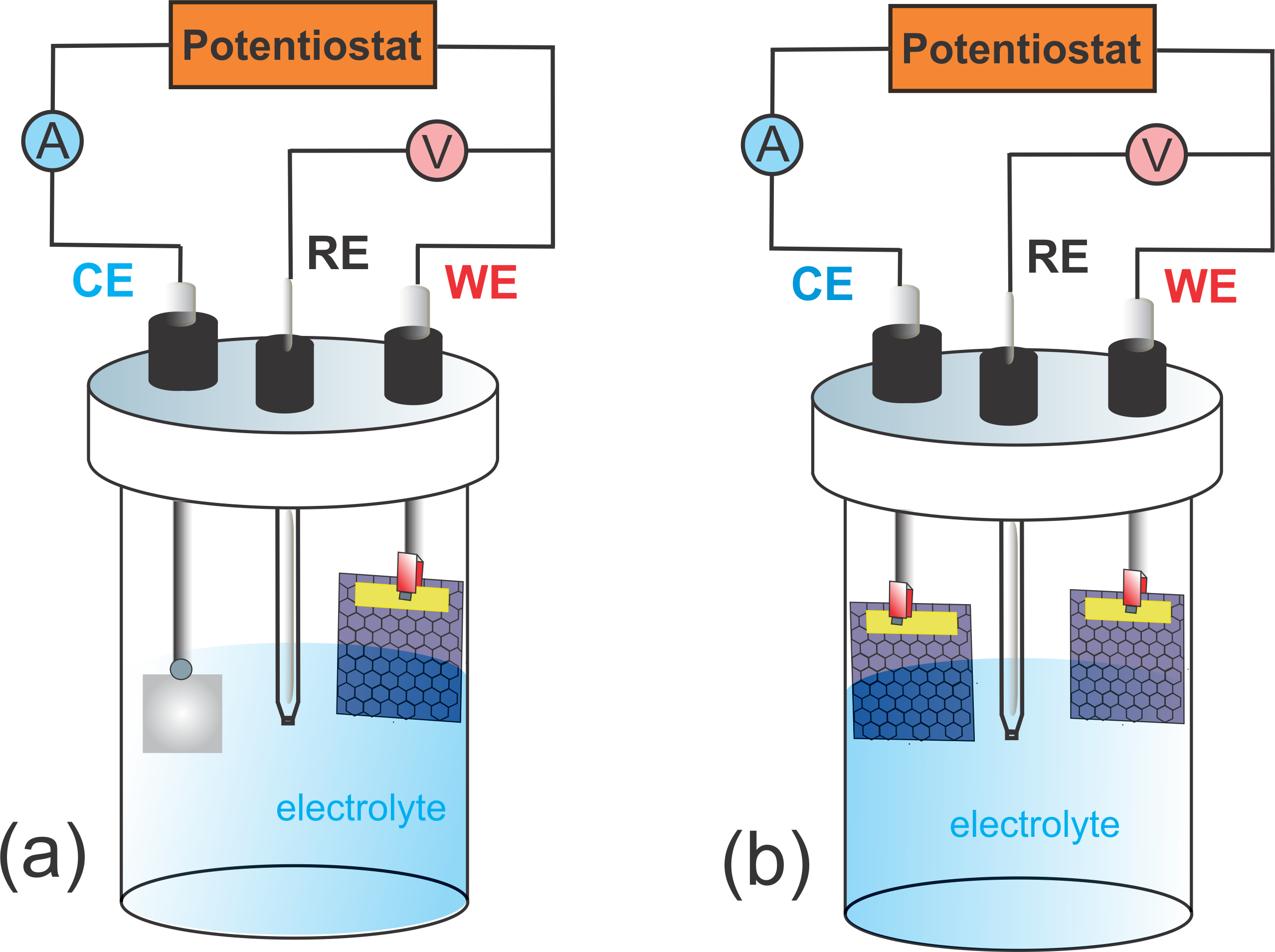}
\caption{Illustration of the experimental setup of a three-electrode electrochemical cell configuration. (a) Cell setup used to measure the quantum RC electrodynamics~\citep{Lopes-2024-SLG-structure,Bueno-2022,lopes2021measuring} of an SLG in which the SLG is used as the WE and Pt foil is used as the CE. (b) Cell setup used to measure and demonstrate the quantum entanglement between two equivalent SLGs possessing equivalent quantum RC electrodynamics.}
\label{fig:entanglement-setup}
\end{figure} 

As noted in Section ~\ref{sec:Operation-Potentiostat}, a potentiostat is electronic instrument for measuring electrochemical processes and operates using a three-terminal setting, where an electric potential perturbation between the WE and RE terminals can be imposed to measure electric currents flowing between the WE and CE. In this three-terminal measurement setting, the impedance of the system in contact with the WE was independent of the impedance in contact with the CE. Accordingly, the chemical or electric changes occurring in the CE did not affect the measurements taken in the WE.

This three-terminal operational setting of a potentiostat allows us to study quantum electrochemistry by measuring the circuit impedance of molecular or nanometric scale moieties in contact with the WE of the electrochemical cell, in which the prevailing circuit response is a resistance quantum placed in series to the quantum capacitance of the systems~\citep{Pinzon-2024-Meletronics, Bueno-2024-bioreview, Pinzon-2024-QDs, Lopes-2024-SLG-structure, Bueno-Davis-2020, Garrote-2020-Nature, Bueno-2022, bueno2023quantum, sanchez2022quantum, lopes2021measuring, alarcon2021perspective}. This was also the case for the impedance of graphene sheets embedded in an electrolyte environment~\citep{lopes2021measuring, Bueno-2022}, as shown in Figure~\ref{fig:entanglement-setup}\textit{a}, because the impedance of the CE resulted from the Pt foil. This is the case when the CE's impedance is a classical circuit element, in which the CE terminal of the potentiost is in contact with a Pt foil (with a negligible $C_q$ owing to its very high DOS), which serves as the classical CE terminal of an electrochemical cell. However, this does not occur if the impedances of the WE and CE are those of quantum RC circuit elements.

As confirmed in Section ~\ref{sec:Operation-Potentiostat}, the $Z_{CE-RE}$ impedance placed in the CE terminal can be equal to or different from that measured in the WE terminal and does not affect the impedance measurements in the WE terminal of the potentiostat if a three-terminal setting is used. By considering this electronic engineering of potentiostat equipment, further investigations were conducted in which quantum RC circuit elements within a graphene sheet in a time-dependent measurement were poised in contact with both the WE and CE of an electrochemical cell operating in a three-terminal setting.

Graphene sheets exhibit quantum electrodynamics that obey Eq. ~\ref{eq:Complex-Cq}, as previously demonstrated in the literature ~\citep{lopes2021measuring,Bueno-2022}. The impedance of a graphene sheet is measured as the quantum resistance $R_q$ in series with the quantum capacitance $C_q$ of the sheet, thus following quantum RC dynamics~\citep{lopes2021measuring,Bueno-2022}. Although the relaxation dynamics are mathematically similar to those of a classical series RC circuit, the experimental quantitative analysis confirms that the resistance corresponds to $h/e^2$ whereas $C_q$ is directly associated with the density of states of the system. This finding allowed us to determine the electronic structure of graphene using electrochemical measurements~\citep{Lopes-2024-SLG-structure}.

Accordingly, to measure the impedance of the SLG, the SLG was placed in contact with the WE of the potentiostat and the CE terminal was placed in contact with a Pt foil, as shown in Figure~\ref{fig:entanglement-setup}\textit{a}.  Conversely, in the case in which another equivalent SLG is connected to the CE instead of the Pt foil, as depicted in Figure~\ref{fig:entanglement-setup}\textit{b}, a non-classical correlation between the WE and CE is established, in which the capacitance of the cell is approximately half of that measured in the case depicted in Figure~\ref{fig:Nyquist-comparison}\textit{a}. This implies non-classical correlations between the WE and CE of the potentiostat, as demonstrated previously, which is attributed to a quantum entanglement phenomenon consistent with the Hilbertian subspace mathematical examination of the experiment.

In the next section, both setups depicted in Figure~\ref{fig:entanglement-setup}, that is, the measurement of an SLG solely in contact with the WE or the quantum mechanical impedance of the WE entangled with a second SLG foil in contact with the CE, will be quantitatively discussed.

\begin{figure}[t]
\centering
\includegraphics[width=8cm]{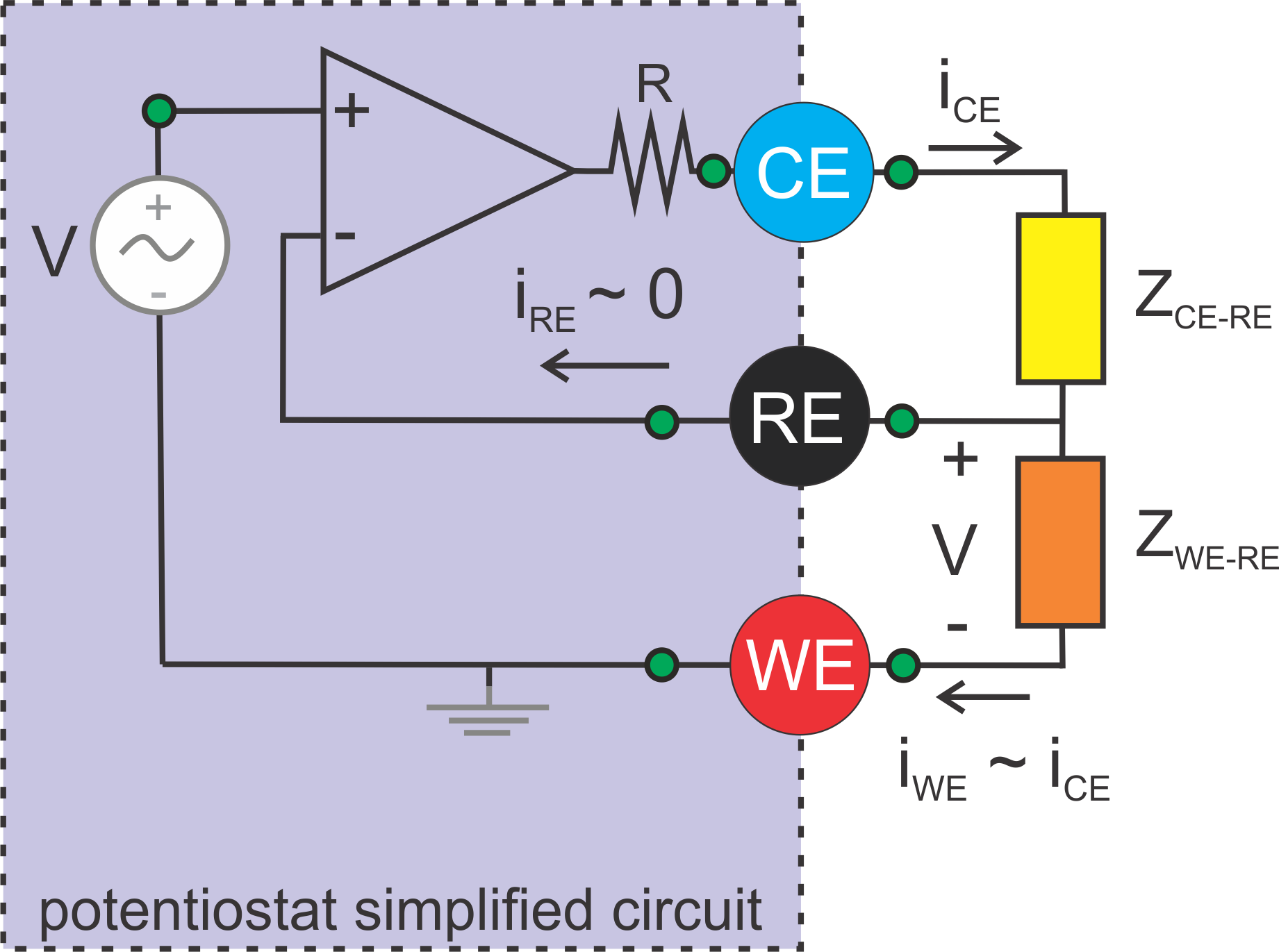}
\caption{Simplified circuit of a potentiostat operating in a three-terminal electrochemical setting in which the potentiostat acts as an operational amplifier (shown by the dotted lines). Note that in this system, the RE acts as a nonpolarizable reference electrode that imposes a potential perturbation to the WE. The electric circuit is closed between the WE and CE and an electric current cannot flow from the CE terminal to the operational amplifier because of the perturbation of the WE via the RE. The operation of this type of potentiostat configuration was investigated considering two electrical impedances: $Z_{WE-RE}$ and $Z_{CE-RE}$, as indicated in Figure~\ref{fig:potentiostate-behavior-setups}\textit{b}.}
\label{fig:three-electrode-setup}
\end{figure}

\subsection{Entanglement between Two Electronic and Spatially Separated Graphene Sheets}
\label{sec:experimental-non-classical-correlations}

\begin{figure*}[t]
\centering
\includegraphics[width=18cm]{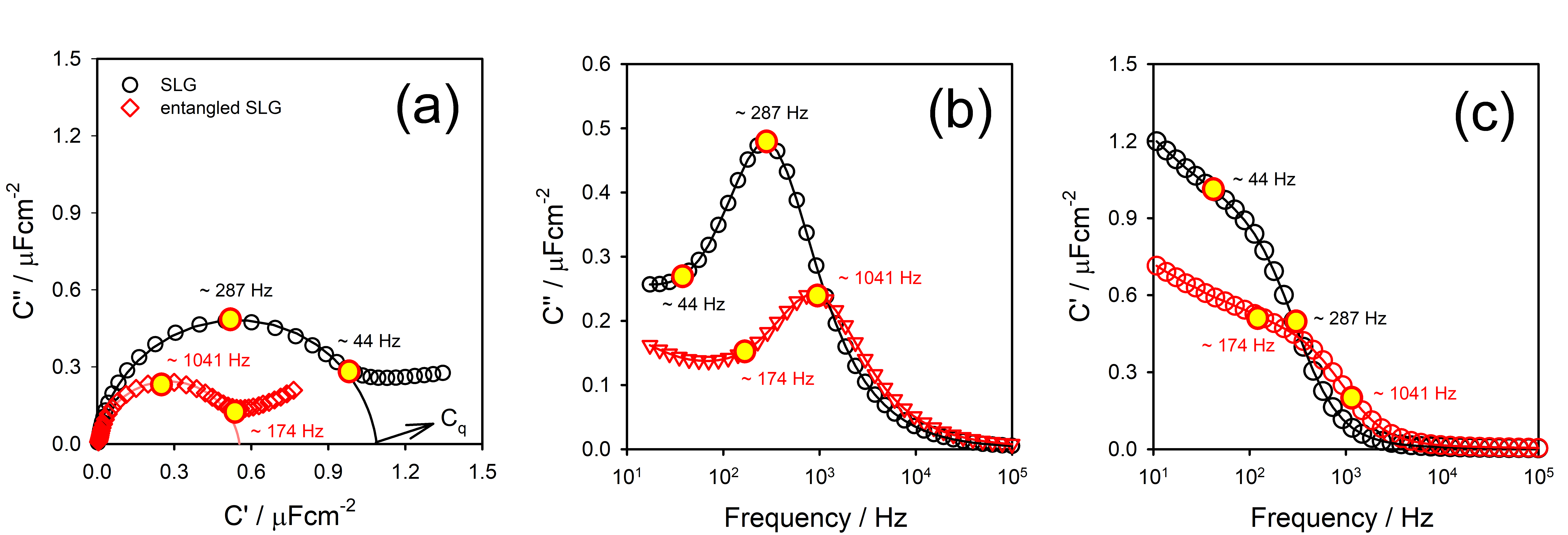}
\caption{Impedance measurements were conducted in a three-terminal potentiostat configuration at the Dirac point ($-0.13 V$) as the stationary energy state of the system for the single (denoted as SLG, black opened circles) and entangled (denoted as entangled SLG, red opened circles) graphene sheets. Here, “single” refers to the measurement setting in which an SLG was contacted to the WE and Pt foil was in contact with the CE. In contrast, “entangled” refers to the setting in which both WE and CE were contacted to equivalent SLGs. (a) Nyquist capacitive plots. (b) Imaginary and (c) real capacitive Bode plots. The lines are mathematical adjustments to the quantum dynamics predicted by Eq.~\ref{eq:Complex-Cq}. More details of the results of the fitting and quantitative equivalent circuit analysis are provided in Section S2.}
\label{fig:Nyquist-comparison}
\end{figure*}

\begin{figure}[t]
\centering
\includegraphics[width=8cm]{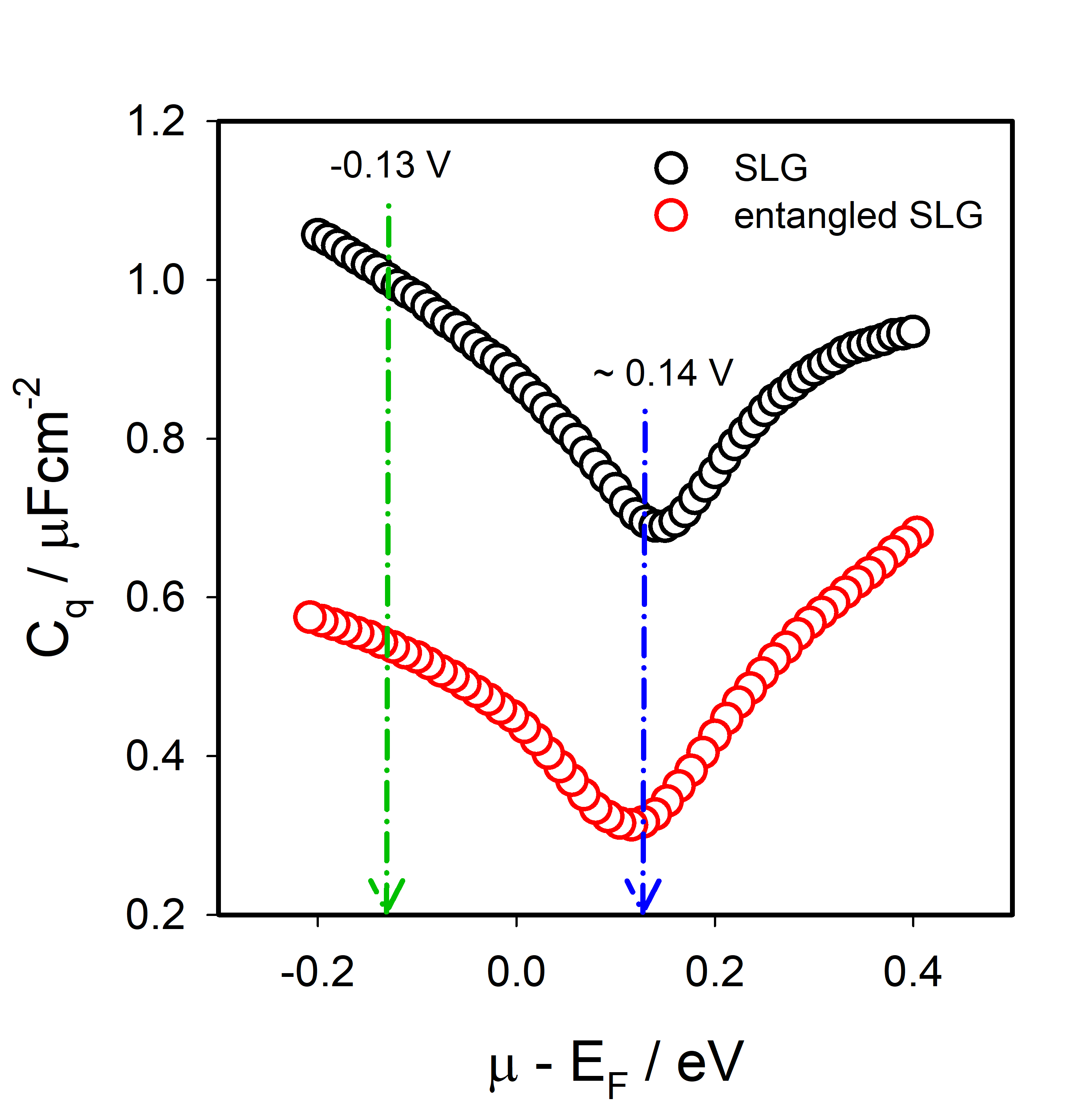}
\caption{Capacitive V-shapes measured in PBS electrolyte as described in reference~\citep{lopes2021measuring} for the single and entangled SLG settings depicted in Figure~\ref{fig:entanglement-setup}. The values of OCP $\sim$ -0.13 V (green arrow) and of the Dirac point $\sim$ 0.14 V (blue arrow) measured in the Ag$\mid$AgCl chemical potential reference are indicated. Within the experimental errors of the measurement within quadruplicates, the Dirac point is equivalent in the single and entangled settings.}
\label{fig:V-shapes}
\end{figure}

Note that in both settings, that is, with a single SLG placed in the cell or entangled with another SLG in the CE, each complex capacitance response of the WE terminal of the cell, as depicted in Figure~\ref{fig:Nyquist-comparison} (the plots show the average value of a quadruplicate measurement), corresponds to the impedance measurements of quantum RC circuits. This is also evidenced from the fitting of the data to Eq. ~\ref{eq:Complex-Cq}, as quantitatively depicted in Table S2 of the SI document. These findings imply that both responses correspond to quantum RC circuits within quantum electrodynamics, obeying QR theory with equivalent Dirac points (as shown in Figure~\ref{fig:V-shapes}) and quantum resistance values (with calculations and quantification demonstrated in Section SI.2 ).

When the Pt foil is used as the CE, no influence of the impedance of the Pt foil on that measured in the WE is observed, and the quantum RC dynamics solely result from the impedance of a single SLG foil contact with the WE of the electrochemical cell. Conversely, when the CE terminal of the cell was connected to a second and equivalent SLG foil, the impedance measured in the WE was affected by the impedance of the other SLG in contact with the CE. This is an unexpected non-classical effect that was interpreted as a quantum entanglement between the two graphene sheets . Quantitatively, the capacitance of the entangled system is approximately half that without entanglement, as the capacitance of the entangled system corresponds to the series combination of the equivalent-magnitude quantum capacitance of two equivalent SLGs. Furthermore, the frequency of the entangled system was approximately four times that in the absence of the entanglement phenomenon. The quantitative analysis obtained from the fitting of Eq. ~\ref{eq:Complex-Cq} with the data shown in Figure~\ref{fig:Nyquist-comparison} agrees with the QR theoretical analysis of the observed entanglement phenomenon (see Section SI.2 for more detail).

The energy of a single graphene foil~\citep{Lopes-2024-SLG-structure,lopes2021measuring} was measured as $E = g_s g_e(e^2/C_q)$, where $g_e$ corresponds to the energy degeneracy of the electrostatic and quantum capacitances imposed by the electrolyte screening dynamics~\citep{Lopes-2024-SLG-structure}. The frequency of the graphene foils measured in an electrolyte environment is given by $\nu = E/h = g_s g_e (e^2/hC_q)$~\citep{Lopes-2024-SLG-structure,lopes2021measuring}, where $g_s$ is the extra degeneracy contribution owing to the spin degeneracy of the electron. This implies a conductance quantum that is two-times the reciprocal of von Klitzing constant~\citep{klitzing1980new} $R_K = h/e^2 \sim$ 25.8 k$\Omega$, corresponding to $g_s e^2/h \sim$ 77.5 $\mu$S. This analysis is in quantitative agreement with the impedance responses depicted in Figure~\ref{fig:Nyquist-comparison}, which show that the entangled system has half the capacitance of the non-entangled system, corresponding to double the $E = g_e e^2/C_q$ value and quadruple the $\nu = g_s (g_e e^2/hC_q)$ frequency. Having the capacitance corresponds to doubling the contribution of the total energy by doubling the energy by doubling the number of foils. There is an additional contribution of $g_s$ to the frequency of the entangled system, with an additional spin dynamics contribution. Thus, $\nu$ of a single SLG is quadrupled in the presence of two entangled SLGs with extra contributions owing to $g_s$ and $g_e$ degeneracy states.

To summarise, the energy doubled and frequency quadrupled owing to the contributions of $g_s$ and $g_e$ associated with the $b$ graphene foil subsystem entangled with the original $a$ system. This formed an entangled $a,b$ system, as quantitatively explained via the QR interpretation of the observed entanglement phenomenon.

This entanglement phenomenon associated with quantum electrodynamics, owing to the entanglement between quantum electric circuits, is novel. Entanglement between quantum circuits is associated with the time-dependent perturbation of fermionic particles within each quantum circuit subsystem. To emphasise the time-dependent characteristics of this measurable phenomenon, Figure~\ref{fig:three-electrode-setup} presents a simplified schematic circuit visualisation of the electronics of a potentiostat within an FRA module. This figure indicates that a potentiostat operating in a three-terminal electrode configuration can measure the impedance (or perform time-dependent measurements) by imposing an alternating voltage perturbation $V(t) = \bar{V} + V_0 \exp(j \omega t)$ in the WE with respect to the RE. This allows the measurement of an alternating electric current response $i(t) = \bar{i} + I_0 \exp(j\omega t - \phi)$ of the CE concerning the WE 

In this experimental measurement that considers entanglement between two quantum circuits with equivalent characteristic time constants and similar frequencies, there is no electron particle electric current flowing between the CE and WE as the measurement occurs under equilibrium conditions. Under such conditions, the net stationary DC-type potential difference, established between the WE and CE, is null. 

Therefore, there was no direct electric current between the WE and CE. The measured time-dependent electric current $i(t)$ response in the potentiostat is a displacement electric current, which is monitored by time-dependent (AC-type) measurements but not by DC-type measurements. The time-dependent nature of the measurements ensures that the displacement electric current flowing through $Z_{CE-RE}$ and $Z_{WE-RE}$ is equivalent, because $Z_{CE-RE}$ and $Z_{WE-RE}$ are equivalent quantum RC circuits. Classical analysis predicted that the impedance measured in the WE by the potentiostat would solely be $Z_{WE-RE} = (V_0/I_0)\exp(j\phi)$, but the measured entanglement phenomenon showed that $Z_{WE-RE}$ doubles. The displacement electric current $i_0 = C_q (dV/dt)$ and $C_q$ are half that of the non-entangled circuit. Therefore, the doubling of $Z_{WE-RE}$ is may be quantified using QR theory, although this may not be achieved using a classical perspective that does not apply to this system.   

The following section provides further evidence of the entanglement between the two graphene sheets by chemically modifying the SLG in contact with the CE using a biological receptor. The purpose of this modification of the SLG in contact with the CE electrode of the entangled electrochemical cell is to demonstrate the ability to sense the corresponding biological target via the receptor contained in the CE by measuring the impedance of the WE. This would otherwise not be possible in a classical circuit setting in which, for instance, the CE is a Pt foil and its impedance does not affect (or is not entangled to) that of the WE.

\subsection{ Quantum Mechanical Entanglement of the Counter Electrode to the Working Electrode for Sensing Purposes}
\label{sec:covid19-sensing}

In the classical electrodynamic analysis of electrochemical events measured in a three-terminal setting, the impedance of the CE is independent of that of the WE, and any event occurring in the CE cannot be measured or interfered with by the impedance measured in the WE. Accordingly, when sensing a molecule using chemical modifications of the CE, no such influence would be observed in the WE.

\begin{figure}[h]
\centering
\includegraphics[width=8cm]{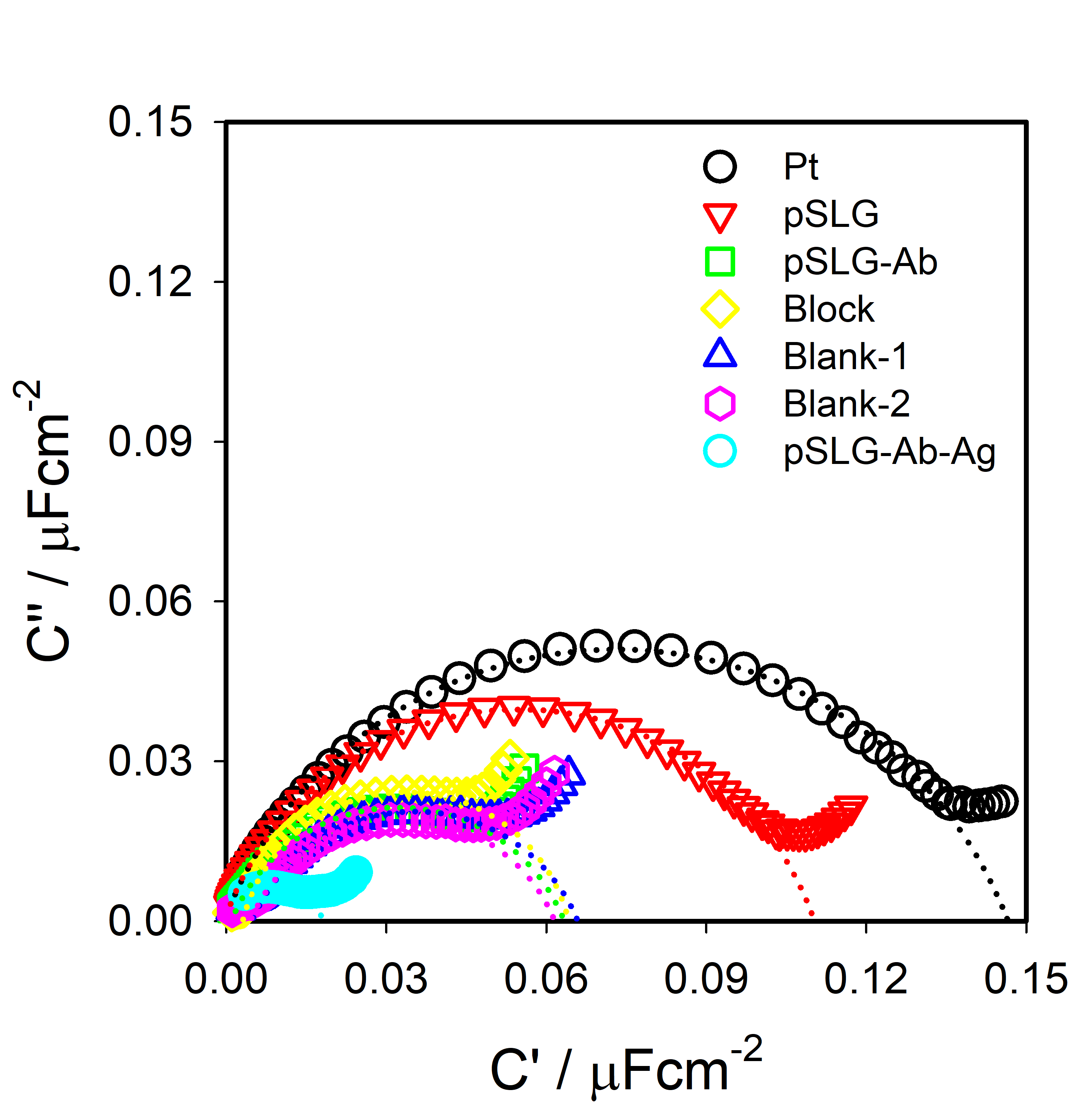}
\caption{Nyquists plots were obtained for an SLG non-covalently modified with 1-pyrene butyric acid in contact with the WE and four different systems contacted to the CE of a three-terminal measurement: platinum (Pt), SLG non-covalently modified with 1-pyrene butyric acid (pSLG), pSLG modified with the antibody (pSLG-Ab), and pSLG incubated with an antigen (Ag) nucleocapsid-protein of SARS-Cov-2 (pSLG-Ab-Ag). The lines correspond to the fitting of the data to Eq.~\ref{eq:Complex-Cq}. More details and the quantitative analysis of these data are provided in Section S3. Measurements referred to as Blank-1 and Blank-2 were conducted after incubating the pSLG-Ab surface in a PBS solution without the presence of the target protein to observe the effect of the environment over the response of the interface without the influence of the Ab-Ag binding event. The response of both blanks is the same, thereby confirming the achievement of biological sensing through an entanglement phenomenon.}
\label{fig:Nyquist-QTeleportation}
\end{figure}

To further investigate if the above traditional electrochemical measurement settings prevail in the case in which the WE and CE are entangled to sense biological events and confirm if the findings of the previous section are valid, a biological sensing assay was conducted. Here, the CE comprised an SLG chemically modified with a biological receptor (antibody) to detect a specific target (nucleocapsid protein, N-protein, of Covid 19). The Nyquist capacitive response of the measurement is shown in Figure~\ref{fig:Nyquist-QTeleportation}, which confirms sensing of the target through the receptor contained in the graphene sheet of the CE. The presence of the target was detected by measuring the impedance of the WE in contact with the SLG that was not incubated in the solution containing the target, demonstrating the possibility of sensing the target by chemical modifications (incubation) imposed in the CE.

The CE sensing experiment provides proof of the entanglement phenomenon, indicating that it  is non-classical and that biological sensing is achieved in the WE, even if the system is not incubated in the solution containing the target species. The WE senses the binding of the target to the receptor in the CE because the WE and CE are entangled. This phenomenon has no classical mechanical analogy, in terms of neither the equivalent classical circuit analysis nor based on the common electronic sense of the operation of a potentiostat (see the analysis in Section ~\ref{sec:experimental_results}).

The next section demonstrates that the entanglement experimentally observed between WE and CE is consistent with the quantum mechanical mathematical analysis of non-classical correlations between a bipartite system. This adheres to what is referred to in particle physics terminology as a quantum entanglement phenomenon.

\section{A Hilbertian Subspace Mathematical Examination of the Results}
\label{sec:theoretical_interpretation}

The preceding sections demonstrated that the operation of a three-terminal electrochemical cell can differ significantly from the classical and traditional behaviour whenever the WE and CE are both comprised of an SLG foil. Whenever this system is time-dependently perturbed, it follows quantum RC circuit dynamics, providing evidence for a non-classical correlation referred to as the quantum entanglement phenomenon. 

This section demonstrates that the preceding measured phenomenon is consistent with the Hibertian subspace mathematical description of an entanglement event.

\subsection{Postulates, Definitions, and Lemmas}
\label{sec:postulates-definitions-lemmas}

We start by considering a system in a Hilbert space $\mathcal{H}$ comprising $N$ subsystems in Hilbert spaces $\mathcal{H}_n$, where $n = 1, 2, \cdots, N$ identifies a quantum subsystem within $N$ possibilities. Following the principles of quantum-mechanical formalism, $\mathcal{H} = \otimes_{i=1}^{N}H_n = H_1 \otimes H_2 \otimes ... \otimes H_n$, where the tensor product is denoted as $\otimes$ \cite{Horodecki2009}. Assuming pure states, the common state of the entire system is referred to as $|\Psi\rangle$; that is,

\begin{widetext}
\begin{equation}
\label{qe:pure_total_state}
|\Psi\rangle = 
\sum_{\bm{\mathrm{i_1, i_2, \cdots, i_N}}} c_{\bm{\mathrm{i_1, i_2, \cdots, i_N}}} |\bm{\mathrm{i_1}}\rangle \otimes |\bm{\mathrm{i_2}}\rangle \otimes \cdots \otimes |\bm{\mathrm{i_N}}\rangle = \sum_{\bm{\mathrm{i_1, i_2, \cdots, i_N}}} c_{\bm{\mathrm{i_1, i_2, \cdots, i_N}}} |\bm{\mathrm{i_1}}\rangle |\bm{\mathrm{i_2}}\rangle \cdots |\bm{\mathrm{i_N}}\rangle,
\end{equation}
\end{widetext}

\noindent in which the tensor product $|\bm{\mathrm{i_1}}\rangle \otimes |\bm{\mathrm{i_2}}\rangle$ can be expressed in terms of Dirac notation as $|\bm{\mathrm{i_1}}\rangle |\bm{\mathrm{i_2}}\rangle$.

To construct an interpretative model for the experimental observations in the present work that considers a particular case within the generalities required for entanglement phenomena, the following postulates were considered:

\begin{itemize}

\item Postulate 1: A quantum particle in a subsystem is a ket in the corresponding Hilbert space.

\item Postulate 2: Quantum entanglement arises between quantum particles.

\item Postulate 3: The state $|\Psi\rangle$ of the entangled system results from the contributions of separable (non-entangled) and entangled states.

\item Postulate 4: The quantum particle considered in the above-introduced postulates obeys quantum electrodynamics within quantum RC dynamics.

\item Postulate 5: The expected value of the energy $E[\rho] = \langle \Psi | \bm{H} | \Psi \rangle$ provided by the Hamiltonian $\bm{H}$ of the system is defined as $E[\rho] \propto g_e (e^2/C_q)$. This agrees with the definition provided in reference~\citep{Miranda-2016}, in which $E[\rho]$ is the energy functional of the electron density $\rho$ and the energy state of the system may be defined by measuring $C_q$ as the only experimental variable in the system.

\end{itemize}

These postulates form the foundation of the theoretical framework required for interpreting experimental quantum electrochemical observations, which follow the premises of quantum electrodynamics, from a quantum entanglement viewpoint.

\textit{Definition: Entangled System.} A system is considered entangled whenever the contribution of separable states is negligible compared to that of entangled states. The entanglement depends on the coupling between the quantum RC rate constants following $\nu = E/h = e^2/hC_q$ quantum electrodynamics in $a$ and $b$ subsystems. Accordingly, the number of separable or entangled states is given by the quantum-capacitance circuit elements of $a$ and $b$ subsystems, which is the only variable in the system where the $h/e^2$ term of the quantum RC circuit dynamics is constant, referred to as the von Klitzing constant~\citep{klitzing1980new}. Whenever $a$ and $b$ subsystems cannot be quantum capacitively entangled, the system is in a separable state with no measurable entanglement.

\textit{Lemma:} Consider a bipartite system composed of two subsystems with $\bm{H_a}$ and $\bm{H_b}$ Hamiltonians, containing $C_{q,a}$ and $C_{q,b}$ quantum capacitance within $C_{q,a}/e^2$ and $C_{q,b}/e^2$ density-of-states, respectively. If the Hamiltonian of the bipartite system is given by $\bm{H} = \bm{H_a} + \bm{H_b}$ and the $C_q$ circuit element in one subsystem is depreciated compared to the other ($C_{q,a} \ll C_{q,b}$), it is assumed that the entanglement phenomenon cannot be measured or observed.

\textit{Proof of the Lemma:} Let $|\phi_i\rangle$ denote the $i$-th particle in the subsystem $a$ with Hamiltonian $\bm{H_a}$ and $|\varphi_j\rangle$ denote the $j$-th particle in the subsystem $b$. Let $C_{q,a} \sim C_{q,b}$. Entanglement has been postulated to exist between quantum particles with an individual quantum capacitance that defines an individual energy level of the particle as $e^2/C_{q,a}$. If particles in the subsystem $a$ are entangled with particles in the subsystem $b$, we obtain $\nu_a = e^2/C_{q,a} \sim \nu_b = e^2/C_{q,b}$ as $C_{q,a} \sim C_{q,b}$. Therefore, these systems are entangled; however, if $C_{q,a} \ll C_{q,b}$, then these systems are not entangled.  When $C_{q,a} \sim C_{q,b}$, the time-dependent dynamics of each subsystem $\nu_n = e^2/hC_{q,n}$ is governed by entangled inseparable states rather than separable states $C_{q,a} \ll C_{q,b}$. Therefore, the quantum entanglement phenomenon can be observed for cases in which $\nu_a \sim \nu_b$, and the system is said to be in an entangled state in quantum mechanical terminology and resonant in terms of electric circuit analysis. 

The definitions and lemmas above provide a mathematical framework that encompasses the measured quantum electrochemical phenomenon as a quantum entanglement event observed in time-dependent, electrochemically driven experiments. The attained entanglement between two quantum mechanical SLG subsystems is only feasible when there is quantum RC entanglement (in quantum mechanical terminology) or a quantum rate $\nu = E/h = e^2/h C_q$ resonance (in electric circuit terminology) between the two subsystems within the equivalent $C_q$ values of the individual particles. The quantum mechanical connections between the particles in each subsystem are solely determined by the individual and equivalent $C_q$ of each subsystem because $e^2/h \sim$ 38.7 $\mu$S is a universal constant, and the only variable~\citep{Miranda-2016} that determines the quantum electrodynamics of the system is $C_q$.

The following sections demonstrate the implications of $C_q$ as a fundamental quantum circuit element of the measured impedance for interpreting the observations of the entanglement phenomenon from a quantum-rate premise and a quantum electrodynamics viewpoint.

\subsection{Terminals of an Electrochemical Cell as a Bipartite System}

It was experimentally demonstrated in Sections ~\ref{sec:experimental-non-classical-correlations} and~\ref{sec:covid19-sensing} that the WE and CE of an electrochemical cell operating in a three-terminal configuration can be entangled if the equivalent impedance measured as quantum RC circuit elements is connected to both the WE and CE terminals of the cell. Section ~\ref{sec:postulates-definitions-lemmas} demonstrated that the above experimental observations comply with the Hibertian subspace mathematical examination of the problem.

For example, let us denote the WE terminal of the cell as $a$ and the CE terminal as the $b$ subsystem of the bipartite system defined in Section ~\ref{sec:postulates-definitions-lemmas}, where each Hamiltonian of the subsystems is denoted $\bm{H_a}$ and $\bm{H_b}$, respectively. In this depiction of the terminals of the cell as a bipartite system, $|\phi_i\rangle$ represents the $i$th particle in subsystem $a$ and $|\varphi_j\rangle$ is the $j$th particle in subsystem $b$. Therefore, the state $|\Psi\rangle$ of the system is expressed as

\begin{widetext}
\begin{equation}
\label{eq:bipartite_state}
|\Psi\rangle = \sum_i a_i |\phi_i\rangle \sum_j b_k |\varphi_j\rangle + \sum_{i,j} e_{i,j} |\phi_i\rangle |\varphi_j\rangle =
|\phi\rangle |\varphi\rangle + \sum_{i,j} e_{i,j} |\phi_i\rangle |\varphi_j\rangle,
\end{equation}
\end{widetext}

\noindent where the separable contribution is $|\phi\rangle |\varphi\rangle$, the entangled system is $\sum_{\bm{\mathrm{i_1, i_2}}} c_{\bm{\mathrm{i_1, i_2}}} |\bm{\mathrm{i_1}}\rangle |\bm{\mathrm{i_2}}\rangle = \sum_{i,j} e_{i,j} |\phi_i\rangle |\varphi_j\rangle$, and the Hamiltonian that describes the quantum mechanical properties of the entire system can be written simply as $\bm{H} = \bm{H_a} + \bm{H_b}$. 

This can be mathematically demonstrated by analysing  Eq. ~\ref{eq:bipartite_state}, where $\bm{H_a} \neq \bm{H_b}$, which is the case for the experiments conducted in the configuration depicted in Figure~\ref{fig:entanglement-setup}\textit{a}. The WE and CE are separable subsystems because there is no quantum capacitance entangling the WE and CE. Conversely, experiments conducted in the configuration depicted in Figure~\ref{fig:entanglement-setup}\textit{b} imply $\bm{H_a} \thicksim \bm{H_b}$ owing to equivalent quantum RC dynamics where $C_{q,a} \sim C_{q,b}$, resulting in $\nu_a = e^2/hC_{q,a} \sim \nu_b = e^2/hC_{q,b}$ (where \textit{a} is equal to the spatially separated \textit{b} system). According to Eq. ~\ref{eq:bipartite_state}, these systems are entangled, as mathematically demonstrated in Section ~\ref{sec:postulates-definitions-lemmas}.

\subsection{Bell's Interpretation of the EPR Paradox}

In 1964, Bell proposed mathematical interpretations~\cite{Bell1964} of the EPR paradox~\cite{EPR1935}. Drawing inspiration from the Bohm-Aharonov analysis~\cite{Bohm1957}, Bell demonstrated that a separable calculation of the expectation value in an entangled system is impossible if quantum mechanics satisfies both locality and causality. He introduced a mathematical methodology to examine whether quantum mechanics was a complete theory. Experimental studies conducted by Alain Aspect~\cite{Aspect1982a, Aspect1982b} confirmed the invalidity of Bell's inequalities in quantum-entangled systems, indicating nonlocal effects associated with the nature of quantum–mechanical entanglement.

Bell's interpretation of the EPR paradox was consistent with that of  quantum entanglement. To demonstrate this, consider a quantum particle $a$ (denoted as $|\phi_i\rangle$) in the WE subsystem, and another particle $b$ (denoted as $|\varphi_i\rangle$) in the CE subsystem. Both the $a$ and $b$ subsystems form part of a complete quantum mechanical system established by the coupled quantum mechanical impedance of the potentiostat. Bell hypothesised that if there is locality, there is a possibility of a mathematical separation of the expectation value of the quantum mechanical states. Hence, let $\lambda$ represent the hidden variables proposed by Einstein with a probability distribution of $\rho(\lambda)$, and let $A(|\phi_i\rangle, \lambda)$ and $B(|\varphi_i\rangle, \lambda)$ represent the results of the separable measurements performed in the WE and CE, respectively. In Bell's proposal, the expectation value resulting from the observation of $A$ in WE and $B$ in CE is given by $\int d\lambda \rho(\lambda) A(|\phi_i\rangle, \lambda) B(|\varphi_i\rangle, \lambda)$, representing separable measurements with statistical independence (owing to locality) between $A$ and $B$.

In the experimental electrochemical analysis, locality occurs whenever the electrical impedance of the CE does not affect that of the WE, as experimentally noted in Section ~\ref{sec:potentiostate}, where WE ($a$) and CE ($b$) are separable subsystems if CE ($b$) is a classical RC system. The classical RC character of $b$-subsystem prevents entanglement, even though $b$ can be described by the Hamiltonian. Nonetheless, the situation is different whenever there are quantum RC dynamics between particles, with quantum dynamics driving the entanglement between $a$ and $b$ subsystems. For example, whenever an SLG is used as an equivalent quantum impedance element connected to spatially and electrically isolated WE and CE, nonlocality emerges as both systems can entangle, as experimentally demonstrated in Section ~\ref{sec:experimental-non-classical-correlations}. Quantum RC circuit analysis within quantum electrodynamics is compatible with the traditional time-independent quantum entanglement interpretation of the phenomenon.

To summarise, the violation of Bell's inequality occurs whenever the electrical impedance of the WE, measured as a time-dependent potential perturbation of the WE to the RE, is influenced by the time-dependent electric current response of the WE to the CE, in which the displacement electric current is fundamentally critical to the analysis. Therefore, entanglement occurs in the situation depicted in Figure~\ref{fig:potentiostate-behavior-setups}\textit{b}, in which there is an inseparable condition of measurement between the two bipartite systems, and not in that of Figure~\ref{fig:potentiostate-behavior-setups}\textit{a}, where there is a separable impedance measured between the WE and CE terminals of the cell operation in a three-terminal configuration.

\subsection{Quantum RC Electrodynamics Analysis}

A straightforward equivalent circuit analysis of the entanglement phenomenon between two graphene sheets was conducted in this study, which directly connects with the quantum RC electrodynamics premises within a quantum rate analysis that follows the Planck-Einstein electrodynamics, as noted in Section ~\ref{sec:introduction}, such as $\nu = 1/\tau \propto e^2/hC_q$, where $\tau = R_qC_q$. $R_q = h/g_se^2$ is a universal constant that considers $g_s$ as the spin degeneracy of the electron and $C_q$ is correlated to the density of states $(dn/dE) = C_q/e^2$, which for a single and pure state ($dn = 1$) provides $E = e^2/C_q$ as the energy of that state. By considering $eV$ potential perturbations of $E$ imposed by the potentiostat to the individual particles, with an associated electric potential of the quantum states as $V = e/C_q$, the charge state of the WE ($a$) and CE ($b$) can be quantum mechanically entangled as they possess the same energy state level $E_a = e^2/C_q = E_b = \langle\varphi|\langle\phi| \bm{H} |\Psi\rangle$ and the same frequency $\nu_a = e^2/hC_q = \nu_b = (1/h) \langle\varphi|\langle\phi| \bm{H} |\Psi\rangle$.

In agreement with the measured phenomenon, the above analysis implies that time-dependent perturbations $s = dV/dt$ in the electric charges $e = C_{q_a} V$ of the WE ($a$) is measurable as an electric displacement current $i = de/dt = C_{q,b}s$ in the CE ($b$), where $e^2/C_{q,b} =  eV$ is the energy state. Moreover, the impedance of both WE and CE are entangled by a quantum electrodynamics event in which the electric displacement current is important as a phenomenon associated with perturbing the quantized energy state of single charged particle states in a time-dependent manner. This phenomenon can be understood by considering that $C_{q,a}$ and $C_{q,b}$ are electronically connected in a quantum impedance arrangement, and that the charge in the $a$ capacitor is resonant with the charge in $b$. The resonant and adiabatic electric connections can be established at large distances, and the channel connecting the particles is associated with the quantum resistance, which, according to quantum mechanical rules, can occur without scattering (a massless perfect-wave characteristic is allowed for the fermion in graphene~\citep{Novoselov-2005}). Therefore, $R_q \sim h/g_s e^2$ is invariant and adiabatic to the time-dependent potential perturbation conducted in the experiment.

In other words, the time-dependent perturbation of the potential of WE ($a$) to RE ($b$) is measured as an electric displacement current $i = C_q s$ instead of an electric particle current response between WE ($a$) and CE ($b$), where $s = dV/dt = \tilde{V}(t) = V_0\exp(j\omega t)$ is the perturbation between WE ($a$) and RE ($b$) and the entanglement is supposed to occur following to quantum electrodynamics rules in which the meaning of $i = C_q s$ is fundamental. This permits the adiabatic coupling of quantum capacitors through the quantum resistance in which there is massless fermionic quantum electrodynamics and the resting mass  $m_0$ is null. These conditions imply that $E = h\nu = \hbar \textbf{c}_* \cdot \textbf{k}$ is obeyed for the observed quantum entanglement phenomenon, and long-distance communication is possible owing to the quantum electromagnetic characteristic of the phenomenon.

The adiabatic connection between $a$ and $b$ states implies that the entangled $C_{q,a}$ and $C_{q,a}$ capacitive fermionic states are dynamic and intrinsically time-dependent correlated events that last as long as the perturbation is maintained. This mode of entanglement between fermions in $a$ to $b$ subsystems is governed by the density of states $\left( dn/dE \right)$ of each subsystem. For instance, let us consider that $e^2(dn/dE)$ is $C_q$ of the entangled system, such as $g_se^2/C_q = \left( dE/dn \right) = h \left( c_*/\lambda \right)$, where $\lambda$ is the wavelength associated with the wave function of the system. This agrees with the analysis in Section ~\ref{sec:postulates-definitions-lemmas} and provides an interpretation of the entanglement phenomenon that complies with relativistic quantum electrodynamics. 

The frequency $\nu$ is $c_*/\lambda$ and electrons in the two subsystems can communicate as photons do, in which $c_*$ correlates to the characteristic time of $\tau = R_qC_q$, such that $c_* = \lambda/\tau$. This directly demonstrates that $\nu = 1/R_qC_q \propto e^2/hC_q$, obtained as a result of the meaning of $C_q$, has a fermionic $(dn/dE) = C_q/e^2$ character that permits communication between fermions through perfect quantum channels with $e^2/h$ conductance entwined with $C_q$. For instance, by equating the DOS  of a single fermionic quantum channel $g_s \lambda/c_*h$ to the DOS of a stationary particle $C_q/e^2$, we obtain $\nu \propto c_*/\lambda = e^2/hC_q$, demonstrating the certainty of this time-dependent analysis in describing the quantum entanglement phenomenon in electrochemical environments. The entanglement phenomenon between the intertwined DOS of the $a$ and $b$ subsystems can be experimentally measured and interpreted using quantum electrodynamics principles within quantum-rate theory, representing an advancement in the understanding of quantum entanglement phenomena.

\section{Conclusions}

We investigated non-classical correlations in quantum electrochemical experiments by utilizing comparable quantum RC circuits connected to both the WE and CE electrodes terminals of an electrochemical cell operating in a three-terminal setting. A quantum entanglement phenomenon supported by experimental and theoretical quantum mechanical analysis was demonstrated, which enabled the measurement and quantification of the quantum entanglement phenomenon using time-dependent impedance measurements of the electrochemical cells.

These findings advance our understanding of the behaviour of quantum electrochemical systems and present intriguing prospects for future research. The study of this phenomenon not only helps to advance the design and application of quantum electrochemical devices for the construction of quantum gates and the establishment of advanced quantum mechanical circuit analysis within the electrolyte environment, but it also opens up avenues for developing quantum computing applications using electrochemical setups. This interdisciplinary approach exploits the unique characteristics of quantum electrochemical systems. The fusion of quantum mechanics and electrochemistry represents, as demonstrated by the present study and analysis, an exciting opportunity for future practical quantum technologies in the future.

\begin{acknowledgments}
The authors are grateful to the São Paulo State Research Foundation (FAPESP) for the grants 2017/24839-0 (thematic research project) and 2018/24525-9 (scholarship). Professor Bueno also acknowledges the support of the National Council for Scientific and Technological Development (CNPq) under the grant 305582/2023-2. Part of this work was supported by the \textit{Universidad Industrial de Santander} under the grant VIE 4522.
\end{acknowledgments}




\bibliography{biblio}

\end{document}